\documentclass[amsmath,amssymb,11pt]{revtex4}
\usepackage{bbm}
\def\R{\mathbbm{R}}
\def\C{\mathbbm{C}}
\usepackage{amsgen}
\usepackage{amsfonts}
\usepackage{amssymb}
\usepackage{amsbsy}
\usepackage{epsfig}

\begin{document}
\title{Self-interacting scalar field cosmologies: unified exact solutions and symmetries}
\author{T. Charters${}^{\dag}$, J. P. Mimoso${}^{\S}$}

\address{${}^{\dag}{}$ 
Departamento de Engenharia Mec\^anica/\'Area  Cient\' \i fica  de Matem\'atica\\ 
Instituto Superior de Engenharia de Lisboa\\ Rua Conselheiro 
Em\'{\i}dio Navarro, 1, P-1949-014 Lisbon, Portugal\\ 
Centro de Astronomia e Astrof\'{\i}sica da Universidade de Lisboa \\ 
Avenida Professor Gama Pinto 2, P-1649-003 Lisbon, Portugal} 

\address{ ${}^{\S}$ ${}^{\ddag}$
Departamento de F\'{\i}sica, Faculdade de Ci\^encias da Universidade de Lisboa \\ 
Centro de Astronomia e Astrof\'{\i}sica da Universidade de Lisboa \\ 
Avenida Professor Gama Pinto 2, P-1649-003 Lisbon, Portugal 
} 

\address{${}^\dag$tca@cii.fc.ul.pt, 
${}^\S$jpmimoso@cii.fc.ul.pt,
}

\date{\today}

\begin{abstract}
We investigate a mechanism that generates exact solutions of scalar field
cosmologies in a unified way. The procedure investigated here permits to recover
almost all known solutions, and allows one to derive new solutions as well. In
particular, we derive and discuss one novel solution defined in terms of the
Lambert function. The solutions are organised in a classification  which depends
on the choice of a generating function which we have denoted by $x(\phi)$ that
reflects the underlying thermodynamics of the model. We also analyse and discuss
the existence of form-invariance dualities between solutions. A general
way of defining the latter in an appropriate fashion for scalar fields is put
forward. 
\end{abstract}
\maketitle

\section{Introduction}

During approximately the past three decades there has been considerable interest
in finding exact solutions to the well know scalar field equations in a flat
$4$-dimensional Friedman-Robertson-Walker (FRW) space-time. This was triggered
by the inflationary paradigm
 \cite{Guth:1980zm,Olive:1989nu,Bassett:2005xm,Liddle+Lyth:2000} where a 
scalar field, dubbed the inflaton, plays a central role in producing a brief
stage of accelerated expansion. The search for exact
solutions was to a great extent driven by the need to find a scalar field potential which would
successfully convey the inflationary prescription, i.e., which would produce
sufficient inflation, ending with a graceful exit. Another goal was the
establishment of  a simple and concise way of relating the dynamics of the
scalar field with the CMB observational data and, in particular, to test of the
consistency of the slow-roll approximation in the study of generation
of fluctuations in the large $\phi$ field region  \cite{Lidsey:1995np,Liddle:1994dx}.

The search for exact solutions of models with a self-interacting scalar field
has followed diverse strategies. At first the focal point was the obtainment of approximate
solutions for the cases of potentials assumed to be realistic, because of their
particle physics motivation. Examples of potentials falling into this class were
the spontaneous symmetry breaking double-well potential, the $\lambda \phi^4$
potential, the Coleman-Weinberg potential
 \cite{Guth:1980zm,Olive:1989nu,Bassett:2005xm,Liddle+Lyth:2000}. 
Another perspective in the search for exact solutions
emerged in 1985 by means of which one looked for potentials whose solutions had prescribed
properties. Lucchin and Matarrese  \cite{Lucchin:1984yf} showed the relation 
between power-law inflation and the exponential potential, which was subsequently 
studied in further detail by Halliwell  \cite{Halliwell:1986ja}, Burd and Barrow  \cite{Burd:1988ss},  Barrow\cite{Barrow:1988yc}, Ratra and Peebles
 \cite{Ratra:1989uz}. Other works along this line explored some particular
equations of state  \cite{Barrow:1990,Barrow:1990td,Madsen:1992tv} or arbitrary
time dependences of the scale factor  \cite{Ellis:1990wsa}.     

A third approach has been to explore several recipes for the formal
construction of exact solutions. Muslimov \cite{Muslimov:1990be}, Salopek and
Bond  \cite{Salopek+Bond 90}, and J. Lidsey  \cite{Lidsey:1991zp} 
devised a method to obtain exact solutions using the $\phi$ scalar field as the
independent variable. De Ritis et al   \cite{deRitis:1990ba,Capozziello:1994dn}
explored an alternative method based on the Lie symmetries of the equations to
derive some exact solutions.   

A skilful combination of the previous methods has lead various authors to derive
some classes of new
solutions
\cite{Barrow:1993hn,Barrow:1994nt,Easther:1995pc,Barrow:1995xb,Parsons:1995ew,Chimento:1999th}. 

The quest for exact solutions of single field models proceeds at present, indeed, recently
scalar field cosmologies have been considered as an explanation of  dark energy
called  quintessence, as well as a way to produce the exotic states
dubbed phantom matter \cite{Caldwell:1999ew,Aguirregabiria:2004te,Chimento:2004br,Chimento:2005xa}. In the
latter case one envisages the possibility 
that the kinetic energy of the scalar field  be negative. The interest in
scalar field exact solutions also extends, naturally, to higher dimensional
models \cite{Chimento:1998aq} and to models based on modified gravity theories such as, Brans-Dicke or
superstring and others \cite{Barrow:1994nx,Lidsey:1999mc,Barrow:2001iw}.  

However two questions have not  been fully answered: What is the basic property that
allows the equations to be integrated? What are the similarities between all the
known solutions?   

Here we will be concerned with the question of unifying all the previous
solutions in a single, and simple framework. The procedure that we have devised,
and which is based on a novel generating function, unifies all the solutions
under a single criterion. Furthermore our method, which goes one step beyond
earlier attempts to use the $\phi$ field as an independent
variable~\cite{Muslimov:1990be,Salopek+Bond 90,Lidsey:1991zp}, allows us to
derive new solutions (which we illustrate in subsection \ref{A new  exact
  solution}), and most importantly provides a way to classify all the solutions,
and their corresponding behaviours. Indeed not only does it enable one to make a
complete qualitative analysis of the possible asymptotic behaviour that can be
expected from single field, scalar field cosmologies, but it also permits us to
analyse the form invariance dualities that connect diverse solutions and
potentials. 

In this paper we work in the framework of General Relativity (GR) and shall
consider FRW models with general $N$ spatial dimensions, although many of the
applications some of the applications will be given for $N=3$ to make connection
with the literature. First we show how  the equations of motion for a scalar
field $\phi$ in a flat $(N+1)$-dimensional FRW model can be integrated by
quadrature. This 
integration of  the equations of motion is not based on any particular
technique, but is in fact the result of the simple and particular form of the
equations of motion of a scalar field in FRW 
space-time. The equations of motion of a self-interacting scalar field 
contain a very special non-linear dissipative term proportional to the square root of the
energy of the associated mechanical system. This particular feature is
fundamental and underlies the derivation of a large of number of exact
solutions, and the construction of particular methods to obtain each of them
that can be found in the literature. In the present work we shall present an 
unified scheme that generates almost every exact solutions for the problem. We
illustrate how known-solutions are recovered, and, as an example, we shall
obtain a solution which to the best of our knowledge is novel. 

In the light of the method that we present here, we also analyse the
relations that exist between apparently disconnected sets of solutions. In this
regard we extend Chimento's \cite{Chimento:2002gb} and Chimento and Lazkoz's
\cite{Chimento:2003qy} results, and we 
discuss a form-invariance symmetry that maps a solution of a given set of
equations with a given scalar field potential to another solution of  different
set of equations associated with some other scalar 
field potential. This completes the unification of the set of exact solutions of
the flat Friedman models with a single scalar field. 

\section{Equations of motion}

Let us consider a $(N+1)$-dimensional homogeneous and spatially flat spacetime 
\begin{eqnarray}
  \label{eq:Nfrw}
  ds^2=-dt^2+a^2(t)\sum_{i=1}^{N}(dx^i)^2,
\end{eqnarray}
where $a(t)$ is the scale factor. Assume that the matter content is a scalar
field $\phi$ which is minimally coupled. The Einstein equations can be written
as
\begin{eqnarray}
  \label{eq:einstein1}
  \ddot \phi + NH\dot \phi +  V_{,\phi}=0,\\
  \label{eq:einstein2}
  \frac{1}{2}N(N-1)H^2=  \frac{1}{2}\dot \phi^2+V(\phi),
\end{eqnarray}
where $H=\dot a/a$, with an  overdot representing the derivative with respect to
time. Combining these two
equations one obtain   
\begin{eqnarray}
  \label{eq:H}
  \dot H=-\frac{\dot\phi^2}{N-1}.
\end{eqnarray}
Introducing the new time variable $d\tau=H dt$ and the variable
\begin{eqnarray}
  \label{eq:var}
  x=\frac{\dot \phi}{H},
\end{eqnarray}
 we obtain the following planar, autonomous dynamical system
\begin{eqnarray}
  \label{eq:dynsys1}
  x'&=&-\left(\frac{1}{2}N(N-1)-\frac{1}{2}x^2\right)\left(\frac{2}{N-1}x+\frac{V_{,\phi}}{
      V}\right),\\
  \label{eq:dynsys2}
  \phi'&=&x,
\end{eqnarray}
where the prime stand for the derivative with respect to $\tau$, and $V_{,\phi}$ is the derivative of $V(\phi)$ with respect to $\phi$.
We study the solutions of   (\ref{eq:dynsys1}) and    (\ref{eq:dynsys2})
by considering $dx/d\phi$.
As
\begin{eqnarray}
  \label{eq:xphi}
  \frac{dx}{d\phi}=\frac{x'}{\phi'}=-\frac{\left(\frac{1}{2}N(N-1)-\frac{1}{2}x^2\right)\left(\frac{2}{N-1}x+\frac{V_{,\phi}}{V}\right)}{x}
\end{eqnarray}
we can write
\begin{eqnarray}
  \label{eq:fund}
  \frac{V_{,\phi}}{V}=\left[-\frac{2x}{N-1}-\frac{2x x_{,\phi}}{N(N-1)-  x^2}\right].
\end{eqnarray}

Equation (\ref{eq:fund}) can be approached in two alternative ways. On the one
hand, given the potential $V(\phi)$ we can, in principle, solve this equation to obtain
$x(\phi)$, and thus we can subsequently use  (\ref{eq:dynsys2}) to obtain a
solution. But this is only strictly possible for the case of the exponential potential, as discussed below. On the other hand, we can instead arbitrarily choose $x=x(\phi)$, and
in the sequel obtain the corresponding scalar field potential, as well as an
exact solution through the use of (\ref{eq:dynsys2}). In this latter case, the
integration of equation (\ref{eq:fund}) yields 
\begin{eqnarray}
  \label{eq:Vxphi}
  V(\phi)=A\left(\frac{1}{2}N(N-1)-\frac{1}{2}x^2(\phi)\right)e^{\displaystyle-\frac{2}{N-1}\int    x(\phi)d\phi} 
\end{eqnarray}
where $A$ is an integration constant which determines the amplitude of the potential. 
The specification of a particular form of $x(\phi)$ (satisfying $x^2\leq
N(N-1)$) thus gives the explicit solution
\begin{eqnarray}
  \label{eq:time}
   \int d\tau=\int \frac{d\phi}{x(\phi)},
\end{eqnarray}
where one uses equation (\ref{eq:dynsys2}),
and further utilisation of (\ref{eq:einstein2}) yields
\begin{eqnarray}
  \label{eq:Hphi}
  H(\phi)=\pm\sqrt{A}e^{\displaystyle-\frac{1}{N-1}\int    x(\phi)d\phi}. 
\end{eqnarray}
 Notice that we also have
\begin{eqnarray}
  \dot\phi^2=x^2(\phi)e^{\displaystyle-\frac{2}{N-1}\int    x(\phi)d\phi}.
\end{eqnarray}

Returning to time $t$, one gets,
\begin{eqnarray}
  \label{eq:timet}
   \int dt = \int \frac{d\phi}{\pm \sqrt{A}x(\phi)
      \exp\left(-\frac{1}{N-1}\int x(\phi)d\phi\right)},
\end{eqnarray}
which is the final quadrature.

By selecting a suitable $x=x(\phi)$, and hence $H(\phi)$ from equation (\ref{eq:Hphi}), one 
easily  integrates  the equations
for the scalar field model in closed form. We obtain the solution explicitly in $t$ cosmic time, provided that equation (\ref{eq:timet}) is invertible. As we
shall illustrate in the following section the many solutions sparsely found in
the literature result from specific, simple choices of $x (\phi)$. Our result naturally
shows how the set of existing solutions can be extended. 

From the definition (\ref{eq:var}) we see that $x(\phi)$ is related to the usual
barotropic  index $\gamma=(\rho_\phi+p_\phi)/\rho_\phi$, where
\begin{eqnarray}
  \rho_\phi=\frac{\dot\phi^2}{2}+V(\phi) ,\label{rhophi}\\
  p_\phi=\frac{\dot\phi^2}{2}-V(\phi), \label{pphi}\ \
\end{eqnarray}
in the following way 
\begin{equation}
\label{xgamma}
x(\phi) = \pm \sqrt{\frac{N(N-1)}{2}\gamma(\phi)},
\end{equation}
and thus choosing $x(\phi)$ amounts to a specification of the equation of state. 
However, the present formalism clearly reveals that the choice of the potential
$V(\phi)$ does not uniquely determine the equation of state. Indeed, from
equation (\ref{eq:fund}), we realise that the specification of a potential
is associated with families of solutions $x(\phi)$ which will differ through 
integration constants, and hence, given (\ref{xgamma}), will correspond to different 
thermodynamic regimes.   

If, for instance, one considers $x(\phi)=\lambda$ then it follows directly from equation (\ref{eq:Vxphi})
that
$V(\phi)\propto\exp(-2\lambda\phi/(N-1))$ (see also Section \ref{Exact solutions}), and
the exact solution takes the form 
\begin{eqnarray}
  \label{eq:ffsol1}
  \phi(t)&=&\frac{N-1}{\lambda}\ln\left(t-t_0\right),\\ 
 \label{eq:ffsol2}
  H(t)&=&\pm \frac{(N-1)}{\lambda^2}\frac{1}{t-t_0},\\
  \label{eq:ffsol3}
  V(\phi)&=& A \left(\frac{N(N-1)}{2}-\frac{\lambda^2}{2}\right) e^{-\frac{2\lambda}{N-1}\phi} .
\end{eqnarray}
Notice that the particular case where $\lambda =0$ reduces to the case of a cosmological constant, $V(\phi)= AN(N-1)/2\equiv\Lambda$, and we recover the de Sitter solution  \cite{Madsen:1992tv}.

But if one assumes instead, from the start, the
potential to be exponential (\ref{eq:ffsol3}), then  equation (\ref{eq:fund}) is integrable, and we derive
the most general $x(\phi)$ generating function that yields the exponential
potential.  We obtain though $x(\phi)$ in implicit form  
\begin{equation}
  \label{gen_x_exp}
(\beta-x)^{-(\beta\alpha+\lambda)}(\beta+x)^{\beta\alpha-\lambda}(\alpha x-\lambda)^{2\alpha\lambda} =\Gamma_0   \exp\left( \beta^2\alpha^2 -\lambda^2\right)
\end{equation}
where we have defined $\alpha = 2/(N-1)$ and $\beta=\sqrt{N(N-1)}$, and where
$\Gamma_0$ is an arbitrary integration constant. The  latter expression  cannot be
solved to yield an explicit form for $x(\phi)$. 

However, we see that we have only two possible choices regarding the limit behaviours
of $x(\phi)$ for the exponential potential for the asymptotic limit
$\phi\to \infty$. Consider the 
change of variables $\chi=1/\phi$  \cite{Nunes:2000yc}, then the planar system 
(\ref{eq:dynsys1})
and (\ref{eq:dynsys2}) can be recast in the form
\begin{eqnarray}
  \label{eq:dynsys1_infty}
  x'&=&-\left(\frac{1}{2}N(N-1)-\frac{1}{2}x^2\right)\left(\frac{2}{N-1}x-\lambda\right),\\
  \label{eq:dynsys2_infty}
  \chi'&=&-\chi^2 x,
\end{eqnarray}
which shows that the infinity manifold $\chi=0$, or equivalently $\phi=\infty$,
is invariant. When  $x=x_1=\lambda(N-1)/2$ or $x^2=(x_2)^2=N(N-1)$, $x'=0$ also
vanishes, and these two solutions  correspond to the asymptotic, equilibrium
solutions. They give completely different  asymptotic regimes, both dynamical
and thermodynamical. A local stability analysis reveals that on the infinity
manifold, $x_1$ is stable and $x_2$ unstable for $\lambda^2<4N/(N-1)$ (otherwise, they will be respectively unstable and stable).

When $V(\phi)$ is not exponential we should distinguish the equilibrium
solutions of the system into those arising at finite values of $\phi$ and those
associated with the asymptotic limit $\phi\to \infty$.  In the former case,
equation (\ref{eq:dynsys2}) yields $x=0$ as a necessary condition for a
equilibrium point to occur at finite values of $\phi$. Substitution into the
equation (\ref{eq:dynsys1}) gives  
\begin{equation}
x'= -\frac{N(N-1)}{2}\frac{V_{,\phi}}{V},
\end{equation} 
and thus the left-hand side vanishes if and only if $V_{,\phi}=0$, i.e.,
whenever $V(\phi)$ has an extremum, say  $\phi=\phi_0$. A straightforward linear
stability analysis reveals that the equilibrium point is stable (unstable) if
$V(\phi_0)$ is a minimum (respectively, a maximum).  

On the other hand, if the equilibrium point occurs when $\phi\to \infty$, the
introduction of $\chi=1/\phi$, as done before, shows that an asymptotic
behaviour associated with a non-vanishing constant  value of $x$ only happens if
$V(\phi)$ asymptotes to an positive exponential behaviour, i.e., $\frac{2}{N-1}x
\to\lambda$ in equation~(\ref{eq:dynsys2_infty}). In this latter case the
discussion of the stability previously produced for the exponential case applies.   

Still regarding $\phi\to \infty$ case, we also have, of course, the possibility
that the equilibrium point be characterized by $x=0$ when $V_{,\phi} \to 0$, and
hence, $V\to \Lambda$ corresponding to the de Sitter behaviour. 

Looking at equations (\ref{eq:dynsys1}) and (\ref{eq:dynsys2})  it is easy to
see that it is only for the exponential potential that these two equations 
decouple. For an arbitrary potential $V=V(\phi)$ this does not
occur, and thus the condition for decoupling  is given by the potential
in the form (\ref{eq:Vxphi}), for a given $x=x(\phi)$. This leaves us with only
one equation to integrate $\phi'=x(\phi)$. It seems then natural to assume that
it should exist a transformation that maps the free-field solution, namely
$x(\phi)=cte$, to any other solution with a prescribed $x(\phi)$. This is the
nature of the invariance-form transformation that we are going to construct.

For completeness let us comment on the slow-roll approximation which
is a central (if not a starting) assumption of almost every scenario in
inflationary cosmology. From appropriate combinations of the Hubble parameter
and of its derivatives one can define quantities which take small values when
the slow roll regime of the scalar field dynamics holds. These are called
slow-roll parameters, and subsequently the expressions of  relevant quantities
of the model, such as the spectral scalar and tensor indexes, are written as  
expansions in terms of these parameters in the neighbourhood of the slow-roll
regime \cite{Lidsey:1995np}. Typically, however, only 
the first few enter into any expressions of interest. 

In terms of the generating
function $x(\phi)$ that we have introduced,  the first two slow-roll
parameters are
\begin{eqnarray}
  \label{eq:slow-roll_p}
  \epsilon(\phi)&=&\frac{1}{2}x^2(\phi),\\
  \eta(\phi)&=&\frac{1}{2}x^2(\phi)-\frac{d x}{d\phi}(\phi).
\end{eqnarray}
Apart from a constant of proportionality, $\epsilon$ measures the relative
contribution of the field's kinetic energy to its total energy. The quantity
$\eta$, on the other hand, measures the ratio of the field's acceleration
relative to the friction term acting on it due to the expansion of the
universe. The slow-roll approximation applies when these quantities are small in
comparison to unity, that is, when we have both $x^2(\phi)\ll 1$ and $\frac{d x}{d\phi}(\phi)\ll 1$.
This reduces the dynamical system (\ref{eq:dynsys1}) and (\ref{eq:dynsys2}) to
\begin{eqnarray}
  \label{eq:dynsys1_slowroll}
  x'&\simeq&-\frac{1}{2}N(N-1)\left(\frac{2}{N-1}x+\frac{V_{,\phi}}{
      V}\right),\\
  \label{eq:dynsys2_slowroll}
  \phi'&=&x,
\end{eqnarray}
and this gives
\begin{equation}
  \label{eq:xphi_slowroll}
  \frac{dx}{d\phi}\simeq - N - \frac{V_{,\phi}}{xV}  
\end{equation}
and so $\eta \ll 1$ implies that $\frac{V_{,\phi}}{xV} \simeq - N$, i.e.,
locally $V(\phi)\sim \exp{\left[-N \int x(\phi) d\phi\right]}\sim 1
-N \int x(\phi)  d\phi$.

\section{Exact solutions}
\label{Exact solutions}

In what follows  we list some of the most important exact solutions that can be
found in the literature, and show how they arise as particular cases of the
previous scheme by giving the choice of $x(\phi)$, which in turn determines the
form of the potential $V(\phi)$ for 
each case. This illustrates how the present method recovers, and unifies the
many solutions that populate the literature. These solutions rely on  the  fact
that (\ref{eq:timet}) can be inverted, and this together with the fact that in all cases $x(\phi)$ can be cast as a logarithmic derivative, $x(\phi) =g'(\phi)/g(\phi)$ where $g(\phi)$ is  some well-behaved function of $\phi$, are a common feature for all of them.
(Since in this section our goal is to establish the connection between our procedure and the literature on exact solutions, in what follows we restrict to the $N=3$ case, and we do not review the details of the solutions. For the latter details the reader is kindly referred to the quoted references).

In the second subsection below, we also derive a new exact solution, and discuss its main properties. This enables us to illustrate how the generating method introduced in the present work permits to expand the set of known solutions, and is not limited to reproduce them.

\subsection{List of some of  exact solutions}
\subsubsection{Direct solutions}

\begin{enumerate}
\item One of the most important exact solutions  is the well known power-law
  solution which has been derived in many forms and through different methods  \cite{Lucchin:1984yf}. As we have seen in the previous section, this
  solution corresponds to the choice $x(\phi)=\lambda$, where $\lambda$ is a constant, and yields (for the $N=3$ case) 
  \begin{eqnarray}
    \label{eq:Vpl}
    V(\phi)=A\left(3-\lambda^2/2\right)e^{-\lambda\phi},
  \end{eqnarray}
 where $A$ is an arbitrary constant that fixes the height of the potential (in what follows we will  adopt this notation for the constant factor that defines the amplitudes of the various potentials). 
Note that this solution also permits to integrate the scalar perturbations
giving a constant scalar spectral index \cite{Lidsey:1995np,Liddle:1994dx}.

\item The Easther solution  \cite{Easther:1995pc} corresponds to the choice $x(\phi)=-\phi$ and yields
  the potential
  \begin{eqnarray}
    \label{eq:VEasther}
    V(\phi)=A\left(3-\phi^2/2\right)e^{\phi^2/2},
  \end{eqnarray}
where, as explained, $A$ is an arbitrary constant,
and subsequently gives
\begin{eqnarray}
  a(\phi)&=&\frac{\phi_0}{\phi} ,\\ 
  t(\phi)&=&\frac{1}{2\sqrt{A}}\left[\mathrm{Ei}\left(-\frac{\phi_0^2}{4}\right)-\mathrm{Ei}\left(-\frac{\phi^2}{4}\right)\right],
\end{eqnarray}
where $\mathrm{Ei}$ is the exponential integral function~\cite{Abramovitz 1964}. 
This solution has the remarkable feature that it also yields constant scalar  spectral index which is  equal to $3$.

The Easther solution is a particular case of the class of solutions characterized by $x=\lambda\phi$.
In the latter case the solutions are characterized by
\begin{eqnarray}
    \label{eq:VEasther}
    V(\phi)=A\left(3-\lambda^2\phi^2/2\right)e^{\lambda\phi^2/2},
  \end{eqnarray}
and 
\begin{eqnarray}
  a(\phi)&=& a_0 \phi^{1/\lambda} , \\
  t(\phi)&=&\frac{1}{2\lambda\sqrt{A}}\left[\mathrm{Ei}\left(\frac{\lambda\phi_0^2}{4}\right)-\mathrm{Ei}\left(\frac{\lambda\phi^2}{4}\right)\right],
\end{eqnarray}
where $\mathrm{Ei}$ is, once again, the exponential integral function. 

\item The intermediate inflationary solution \cite{Barrow:1990} is given by $x(\phi)=\beta/\phi$, where $\beta$ is a constant, and is one of the most famous solutions. The scalar field potential is given in this case by
  \begin{eqnarray}
    \label{eq:IInf}
    V(\phi)=\frac{16A^2}{(\beta+4)^2}\left(3-\frac{\beta^2}{2\phi^2}\right)\left[\frac{\phi}{(2A\beta)^{1/2}}\right]^{-\beta},
  \end{eqnarray}
and we have 
\begin{eqnarray}
  a(t)&=& \exp\left(A t^f\right)\\
  \phi&=&\left(2A\beta t^f\right)^{1/2},
\end{eqnarray}
where $f$ is a constant such that $0<f<1$, and $\beta=4(f^{-1}-1)$.

\item In  \cite{Barrow:1994nt} one finds the potentials
  \begin{eqnarray}
    \label{eq:Barrow1994_1}
    V_1(\phi)=A^2\lambda^2\left[(3A^2-2)\cosh^2(\phi/A)+2\right]
  \end{eqnarray}
and 
\begin{eqnarray}
  \label{eq:Barrow1994_2}
V_2(\phi)= \frac{1}{12}\lambda^2A^{-2}\phi^2(\phi^2+A^2)(\phi^4A^{-4}+A²-6+2\phi^2)
\end{eqnarray}
which correspond, respectively, to
\begin{eqnarray}
  a_1(t)&=& a_0\left[\sinh (2\lambda t)\right]^{A^2/2}\\
  \phi_1(t)&=& A\ln[\tanh(\lambda t)]\\
  x_1(\phi)&=& -2 A^{-1}\tanh (\phi/A),
\end{eqnarray}
and to
\begin{eqnarray}
  a_2(t)&=& a_0\left[\sinh(2\lambda t)\right]^{A^2/2}\exp[-A^2\coth^2(\lambda  t)/12],\\
  \phi_2(t)&=&A \mathrm{csch}(\lambda t),\\
  x_2(\phi)&=&-\frac{6\phi}{A^2+\phi^2}.
\end{eqnarray}
In the previous epxressions $A$ and $\lambda$ are arbitrary constants. 

\item In  \cite{Barrow:1995xb} one finds a class of solutions with
  \begin{eqnarray}
    \phi(t)&=& A\exp(-\mu t^n),
  \end{eqnarray}
parametrized by $n$ constant, to which we associate the choice
  \begin{eqnarray}
    x(\phi)=-2 \frac{\phi\left(\log(\phi/A)\right)^{1-1/n}}{\int\displaystyle
    \phi\left(\log(\phi/A)\right)^{1-1/n} { d}\phi}   .
  \end{eqnarray}
In Ref.~\cite{Barrow:1993hn} another family of solutions, also parameterized by constant $n$, is displayed such that
\begin{eqnarray}
  \phi(t)=A(\ln t -B)^n   .
\end{eqnarray}
this family of solutions is recovered 
with the choice
\begin{eqnarray}
  x(\phi)=-2\frac{(\phi/A)^{1/n}\exp((\phi/A)^{1/n})}{\int\displaystyle
    (\phi/A)^{1/n}\exp((\phi/A)^{1/n}) { d}\phi}   .
\end{eqnarray}

In both cases the general form of the potentials is rather involved, except for
some simple cases arising form the restriction of $n$ to take some particular
values, and thus we refer the reader to the cited  references for further
details about the  solutions. 

\end{enumerate}

\subsubsection{Pair defined solutions}
There are other solutions for which a special method was constructed. The
following two examples provide a closer look at the procedures that were used.
\begin{enumerate}
\item In  \cite{Kruger:2000nra} the authors make an assumption which amounts in our prescription  to the choice
  \begin{eqnarray}
    x(\phi)= \displaystyle\frac{[1-F^2(\phi)]\beta^2(\beta-1)^2}{4-[1-F^2(\phi)]\beta(\beta-1)}
  \end{eqnarray}
where $\beta$ is a constant, and thus derived the potential
\begin{eqnarray}
\label{eq:V_Kruger}
  V(\phi)=\exp\left(\mp 2\beta \int\sqrt{\frac{F(\phi)-1}{F(\phi)+1}}{ d}\phi\right).
\end{eqnarray}
The method provides an exact solution with the further prescription  of
\begin{eqnarray}
  F(\phi)=\cosh(\lambda\phi),
\end{eqnarray}
where $\lambda$ is a constant, 
then yielding 
\begin{eqnarray}
  V(\phi)=A(1+\cosh(\lambda\phi))^{\mp\beta/\lambda-1}  .
\end{eqnarray}
Putting $\beta/\lambda=2$ and choosing the positive sign in (\ref{eq:V_Kruger})
one gets the solution
\begin{eqnarray}
  \phi(t)=\frac{1}{\lambda}\ln\left[\frac{\exp(\lambda \sqrt{A}t)+1}{\exp(\lambda \sqrt{A}t)-1}\right]  .
\end{eqnarray}

\item Reference  \cite{Schunck:1994yd} provides another special method which is equivalent to choosing
  \begin{eqnarray}
    x(\phi)=\pm \sqrt{\frac{2g(H)}{H^2}}
  \end{eqnarray}
for which one has
\begin{eqnarray}
  V(\phi)=3H^2(\phi)-g(H(\phi)).
\end{eqnarray}
The solutions of  \cite{Schunck:1994yd}
were obtained with the choices
\begin{eqnarray}
  g_1(H)&=& -A H^n,\\
  g_2(H)&=& \pm \frac{4}{C}\sqrt{AC H-H^2}H.
\end{eqnarray}
Notice that the authors also derived the solution given in   \cite{Barrow:1994nt}
by using a specific choice of the function $g(H)$.

\item In  \cite{Chimento:1999th} a special method was given for a
  $(N+1)$-dimensional FRW space-time using certain classes of generating
  functions using
  \begin{eqnarray}
    x(\phi)=(N-1)\frac{G(\phi)}{\int G(\phi)},
  \end{eqnarray}
where $G(\phi)$ is a  function so that $G(\phi)=\alpha  H(\phi)+L(\phi)$, where $H(\phi)$ and $L(\phi)$ are functions specified below, and $\alpha$ is a constant. 

  \begin{enumerate}
  \item For $H$-linear generating functions and for any $N$, we recover the solutions of \cite{Chimento:1999th} with the following choice of $x(\phi)$
    \begin{enumerate}
    \item \begin{eqnarray}
      x(\phi)=-(N-1)\frac{L(\phi)e^{\alpha\phi/(N-1)}}{\displaystyle \int L(\phi)e^{\alpha\phi/(N-1)}}+\alpha
    \end{eqnarray}
    where 
    \begin{eqnarray}
      L(\phi)=D\alpha e^{-\alpha\phi/(N-1)}\sum_{j=m}^m\frac{j}{j+1} e^{-j \alpha\phi/(N-1)},
    \end{eqnarray}
 for the cases $m=n=1$ and $m=n=2$.

\item And also $L(\phi)=c_1\phi+c_2 \phi^2$ when $c_1,c_2\in\R$.

    \end{enumerate}
\item For the method of the multiplicative generating functions of Ref.~\cite{Chimento:1999th}, the choices of $x(\phi)$ become
  \begin{enumerate}
  \item   $G(\phi)=\omega \phi^n$ and thus
    \begin{eqnarray}
    x(\phi)=\frac{\omega \phi^n}{cte+\omega\phi^{n+1}/(n+1)}.
  \end{eqnarray}

  \item $G(\phi)=\omega \phi^n\exp(-\lambda\phi^m)$ and thus
    \begin{eqnarray}
      x(\phi)=\frac{e^{-\lambda  \phi ^m} m (N-1) \left(\lambda  \phi^m\right)^{\frac{n+1}{m}}}{\phi  \Gamma   \left(\frac{n+1}{m},\lambda  \phi ^m\right)}
    \end{eqnarray}
The special cases were considered: $n=-3$, $m=-2$ and $\lambda=1/2$ for $H_0=0$
and $\omega=-2/\sqrt{3}$; $H_0=1/2$ and $\omega=1+2/\sqrt{3}$ and with $N=3$.
\item $G(\phi)=\omega\phi^n(1-\lambda\phi^m)^\mu$ and thus
  \begin{eqnarray}
    x(\phi)=\frac{(n+1) (N-1) \left(1-\lambda  \phi ^m\right)^{\mu}}{\phi  \ _2F_1\left(\frac{n+1}{m},-\mu ;\frac{m+n+1}{m};\lambda  \phi ^m\right)}
  \end{eqnarray}
The special cases were considered: $n=1$, $m=2$ and $\lambda=\mu=\omega=1/2$
with $H_0=1/(24\lambda)$  and with $N=3$.
  \end{enumerate}
  \end{enumerate}

\item There are also exact solutions for  string
motivated models in   \cite{Easther:1993qg}.
The corresponding choices are
  \begin{eqnarray}
    x_1(\phi)&=&2\xi\frac{A-2Be^{-\xi \phi}}{A-Be^{-\xi \phi}},\\
    x_2(\phi)&=& -2\xi\frac{A+2Be^{-\xi \phi}}{A+Be^{-\xi \phi}},\\
    x_3(\phi)&=& 2\xi\frac{A-3Be^{-2\xi \phi}}{A-Be^{-2\xi \phi}} ,
  \end{eqnarray}
where $\xi$, $A$ and $B$ are arbitrary constants, which  yield the following potentials
\begin{eqnarray}
  V_1(\phi)&=&
  A^2(3-2\xi^2)e^{-2\xi\phi}+2AB(4\xi^2-3)e^{-3\xi\phi}+B^2(3-8\xi^2)e^{-4\xi\phi},\\
  V_2(\phi)&=&
  A^2(3-2\xi^2)e^{-2\xi\phi}-2AB(4\xi^2-3)e^{-3\xi\phi}+B^2(3-8\xi^2)e^{-4\xi\phi},\\
  V_3(\phi)&=&
  A^2(3-2\xi^2)e^{-2\xi\phi}+6AB(2\xi^2-1)e^{-4\xi\phi}+3B^2(1-6\xi^2)e^{-6\xi\phi},
\end{eqnarray}
respectively, and the associated  solutions  given by:
\begin{eqnarray}
  t_1(\phi)&=&
  \frac{1}{A\xi^2}\left[\frac{1}{2}\left(e^{-\xi\phi}-e^{-\xi\phi_0}\right)
    +\frac{B}{A}\ln\left(e^{-\xi(\phi-\phi_0)}\frac{A-2Be^{-\xi\phi}}{A-2Be^{-\xi\phi_0}}\right)
  \right],\\
  t_2(\phi)&=&
  \frac{1}{A\xi^2}\left[\frac{1}{2}\left(e^{-\xi\phi}-e^{-\xi\phi_0}\right)
    -\frac{B}{A}\ln\left(e^{-\xi(\phi-\phi_0)}\frac{A+2Be^{-\xi\phi}}{A+2Be^{-\xi\phi_0}}\right)
  \right],\\
  t_3(\phi)&=&
  \frac{1}{2A\xi^2}\left[\left(e^{-\xi\phi}-e^{-\xi\phi_0}\right)
   + \frac{C}{2}\ln\left(\frac{(1-Ce^{-\xi\phi})(1+Ce^{-\xi\phi_0})}{(1-Ce^{-\xi\phi_0})(1+Ce^{-\xi\phi})}\right)
  \right],
\end{eqnarray}
where $C=\sqrt{3B/A}$.
\end{enumerate}

\subsection{A new  exact solution}
\label{A new  exact solution}
We now derive a new exact solution by exploring the functional forms of
(\ref{eq:Hphi}) and (\ref{eq:Vxphi}) uncovered by the properties of the
Lambert function  \cite{Lambert 96}. Since the solution is novel we revert to the $N$-dimensional case, and later, when appropriate, we shall restrict it to the $N=3$ case.

The Lambert function  is defined to be the function satisfying 
\begin{eqnarray}
  \label{eq:Lambert_def}
  W(\phi)e^{W(\phi)}=\phi,
\end{eqnarray}
and it is used in many applications \cite{Lambert 96}.
When $\phi$ is real, for $-1/e\le \phi
<0$ there are two possible real values of $W(\phi)$  \cite{Lambert 96}. We
denote just by  $W(\phi)$ 
the branch satisfying $-1\le W(\phi)$. Differentiating the
defining equation (\ref{eq:Lambert_def}), and solving for $W'$, we obtain the
following expression for the derivative of $W$:
\begin{eqnarray}
  \label{eq:dW}
  W'(\phi)  &=&\frac{1}{(1+W(\phi))\exp(W(\phi))}\\
  \label{eq:dW2}
&=&\frac{W(\phi)}{\phi(1+W(\phi))},\quad \phi\ne 0.\\
\end{eqnarray}
Another useful relation is the 
integral of $W(\phi)$ given by
\begin{eqnarray}
  \label{eq:intW}
  \int W(\phi) d\phi=\phi\left( W(\phi)+\frac{1}{ W(\phi)}-1\right).
\end{eqnarray}

If $x(\phi)$ is chosen such that
\begin{eqnarray}
  \label{eq:xL}
  \int x(\phi) d\phi=-(N-1)W(f(\phi)),
\end{eqnarray}
then, according to  equations (\ref{eq:dW}) and (\ref{eq:xL}), the pair
$(V(\phi),H(\phi))$ that solves the equations of motion is given   
by
\begin{eqnarray}
  V(\phi)&=&A\left[
    \frac{N(N-1)}{2}\frac{f^2(\phi)}{W^2(f(\phi))}-
    \frac{(N-1)^2}{2}\frac{f'^2(\phi)}{(1+W(f(\phi)))^2}
    \right],\\
    H(\phi)&=&\pm \sqrt{A}e^{W(f(\phi))},
\end{eqnarray}
from which we derive a first integral
\begin{eqnarray}
  \left(\frac{1+W(f(\phi))}{f'(\phi)}\right)^2\dot\phi^2=A(N-1)^2.
\end{eqnarray}
Another first integral is the Friedman equation (\ref{eq:einstein2}).

\begin{center}
  \begin{figure}[ht]
    \epsfig{file=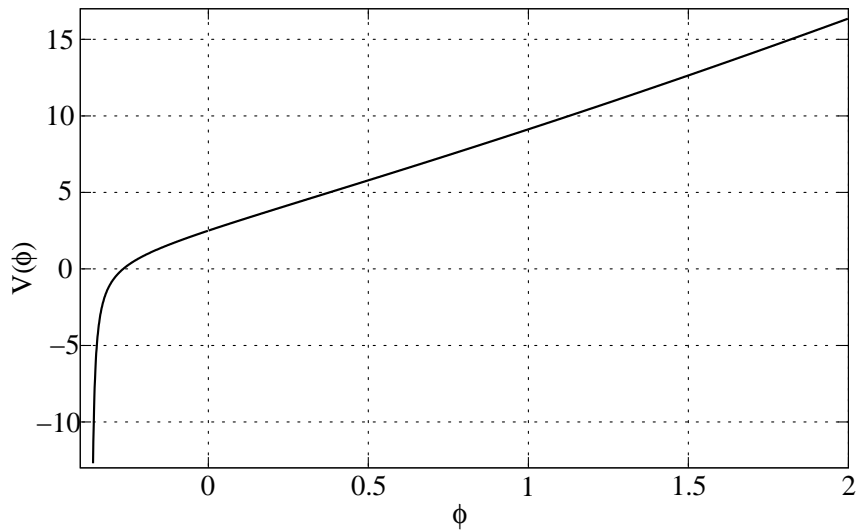, height=7cm}
  \caption{The self-interacting potential as a function of $\phi> -1/e$.}
  \label{fig:VLambert}
\end{figure}
\end{center}

In what follows we consider the $N=3$ case.
In order to find a solution we take $f(\phi)=\phi$ in (\ref{eq:xL}) and in the
following equations. Thus we have $x(\phi)=-2W'(\phi)$ where
$W(\phi)$ is the Lambert function and where $A$ is a constant,
In  this case one gets for the self-interacting potential $V(\phi)$ the
expression 
\begin{eqnarray}
  \label{eq:VLambert}
  V(\phi)=A\left(3-\frac{1}{2}\frac{W(\phi)^2}{\phi^2(1+W(\phi))^2}\right)\frac{\phi^2}{W^2(\phi)}.
\end{eqnarray}
In Figure \ref{fig:VLambert} we show a plot of this potential as a function of
$\phi$. 

The Hubble parameter reads
\begin{eqnarray}
  \label{eq:HLambert}
  H(\phi)=\pm \sqrt{A}e^{W(\phi)},
\end{eqnarray}
and we have plotted $H$ as a function of $\phi$ considering the expanding
branch in (\ref{eq:HLambert}) (the plus sign).

\begin{center}
  \begin{figure}[ht]
    \epsfig{file=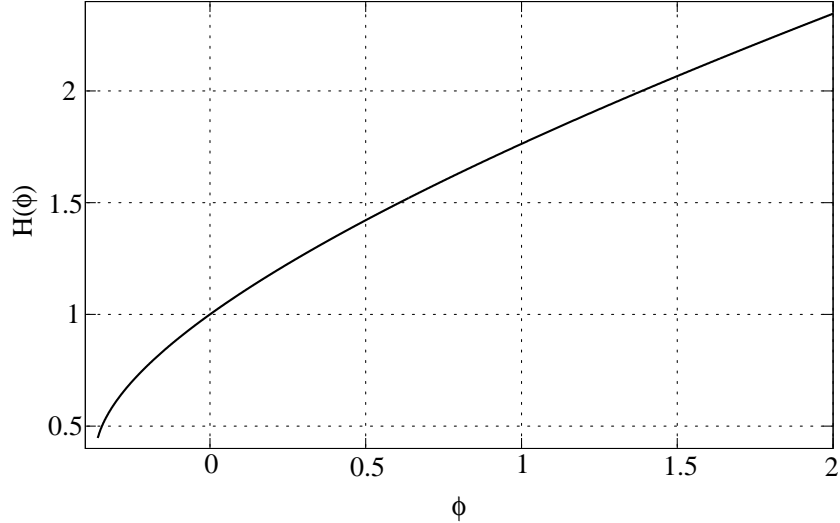, height=7cm}
  \caption{The  Hubble parameter as a   function of $\phi> -1/e$.}
  \label{fig:HLambert}
\end{figure}
\end{center}

Using the definition $x=\dot\phi/H$, it is easy to see that the velocity of the
$\phi$ field is proportional to the condition number of the 
Lambert function, that is, 
\begin{eqnarray}
  \dot \phi=\pm 2\sqrt{A}\frac{\phi W'(\phi)}{W(\phi)}=\pm 2\sqrt{A}\frac{1}{1+W(\phi)},
\end{eqnarray}
and using  (\ref{eq:intW}), one gets,
\begin{eqnarray}
  \label{eq:phi_t}
  \pm 2\sqrt{A}(t-t_0)=\phi\left(\frac{1}{W(\phi)} + W(\phi)\right).
\end{eqnarray}
These equations cannot be easily inverted, but two different asymptotic solutions
can be obtained. If $\phi \ll e$ then (\ref{eq:phi_t}) yields
\begin{eqnarray}
  \label{eq:L_sol_1}
  \phi(t)=\pm 2 \sqrt{A}(t-t_0)\ln\left(\pm 2\sqrt{A}(t-t_0)\right).
\end{eqnarray}
For $\phi \gg e$ one gets
\begin{eqnarray}
  \label{eq:L_sol_2}
  \phi(t)= 2 \sqrt{A}\frac{t-t_0}{2 W(\pm \sqrt{2}/2 \sqrt[4]{A}(t-t_0))}.
\end{eqnarray}

Because
\begin{eqnarray}
  \label{eq:Wass}
  W(\phi)\sim \log \phi -\log(\log(\phi)),\quad \phi \gg 1
\end{eqnarray}
the asymptotic form of the potential for large values of $\phi$ is given by
\begin{eqnarray}
  \label{eq:Vass}
  V(\phi)\sim A \left[\frac{3 \phi^2}{\log(\phi/\log\phi)}-\frac{1}{2\left(1+\log(\phi/\log\phi)\right)^2}\right],
\end{eqnarray}
which is thus the form of the potential $V$ for the slow-roll regime.

The slow-roll parameters  \cite{Stewart:1993bc,Liddle:1994dx,Lidsey:1995np} read
\begin{eqnarray}
  \label{eq:sr_Lambert1}
  \epsilon&=&2\frac{W^2(\phi)}{\phi^2(1+W(\phi))^2},\\
  \label{eq:sr_Lambert2}
  \eta &=& 2\frac{W^2(\phi)}{\phi^2(1+W(\phi))^2}-2\frac{e^{-2W(\phi)}(2+W(\phi))}{(1+W(\phi))^3}.
\end{eqnarray}
\begin{center}
  \begin{figure}[ht]
\epsfig{file=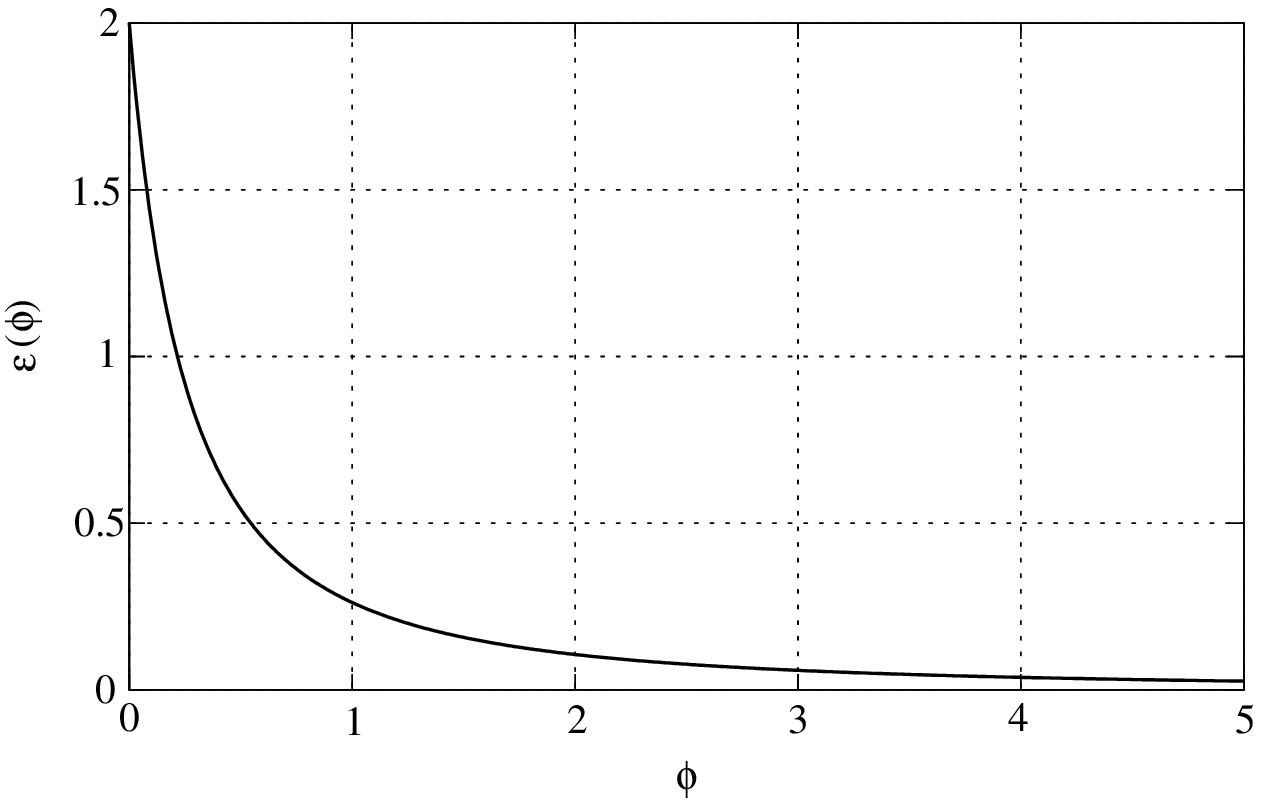, height=7cm}

\epsfig{file=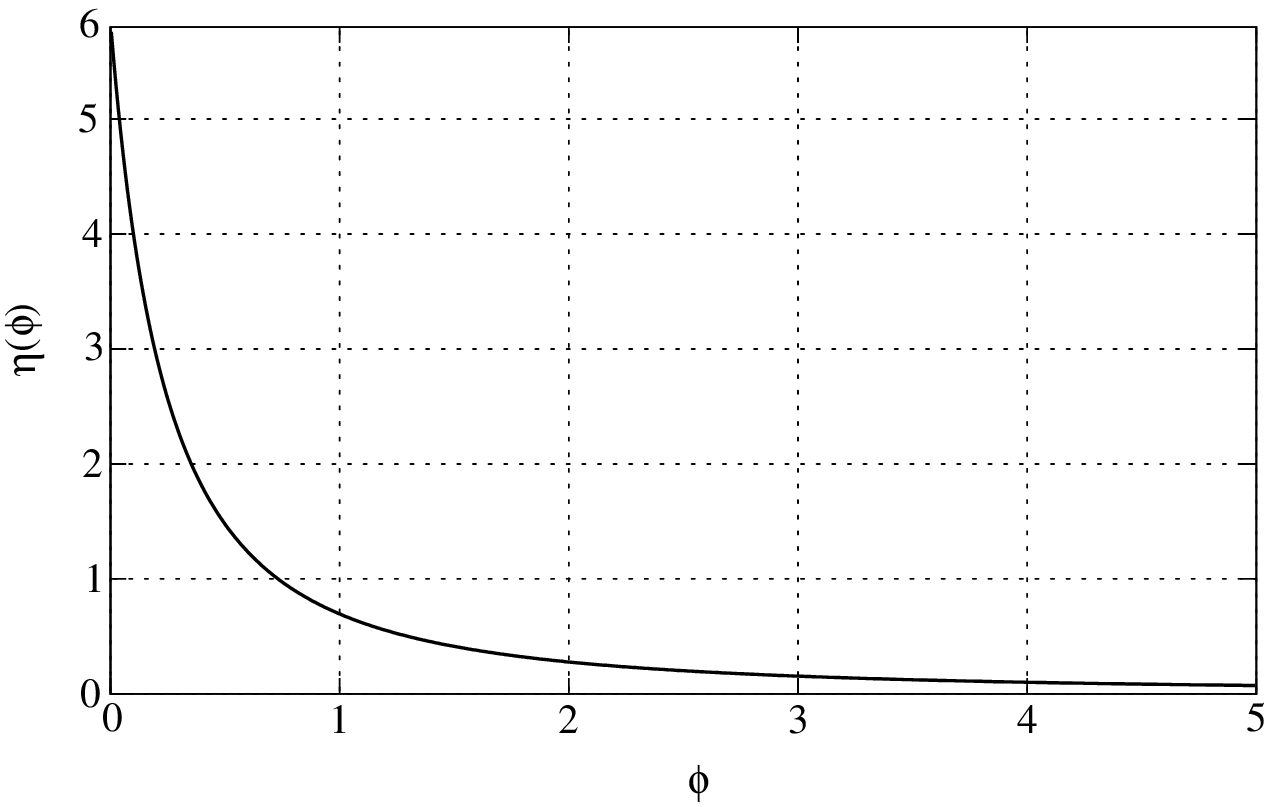, height=7cm}
  \caption{The first two slow-roll parameters as a function of $\phi$.}
  \label{fig:SRVLambert}
\end{figure}
\end{center}
Notice that $\epsilon\le 2$ for $\phi>0$ and that $\epsilon\to 0$ as $\phi\to
+\infty$.  In Figure \ref{fig:SRVLambert}  we show the values of the first and second slow-roll
parameters as a function of $\phi$, notice that both converge to zero when
$\phi$ goes to infinity.  So we see that slow-roll inflation takes place at large values of $\phi$. At  $\phi=0$, i.e., for small values of $\phi$, the model has a radiation-like behaviour (since $\epsilon = 3\gamma/2$, and hence $\gamma=4/3$ implies $\epsilon=2$).

We recall that the condition for inflation to occur is $\epsilon < 1$ and that it ends
at $\epsilon= 1$. If we assume that the scalar field has a large initial value 
the inflationary period ends for the value of $\phi^*$ that solves the
non-linear equation, see (\ref{eq:sr_Lambert1}),
\begin{eqnarray}
  \label{eq:epsW}
  W'(\phi)=\frac{\sqrt{2}}{2}.
\end{eqnarray}
It turns out that in this case the value of $\phi$ that corresponds to the end
of inflation can be determined explicitly. This is one of the most remarkable
properties of the Lambert function. In order to obtain $\phi^*$ consider
equation (\ref{eq:dW2}). Invert, and then multiply both sides by $e$ and notice
that 
\begin{eqnarray}
  1+W(\phi)=W(e\sqrt{2}),
\end{eqnarray}
using $\phi=W(\phi)e^{W(\phi)}$ one gets
\begin{eqnarray}
  \label{eq:phi_endINF}
  \phi^* =\sqrt{2}\left(1-\frac{1}{W(e\sqrt{2})}\right)\simeq 0.21631
\end{eqnarray}

\begin{center}
  \begin{figure}[ht]
\epsfig{file=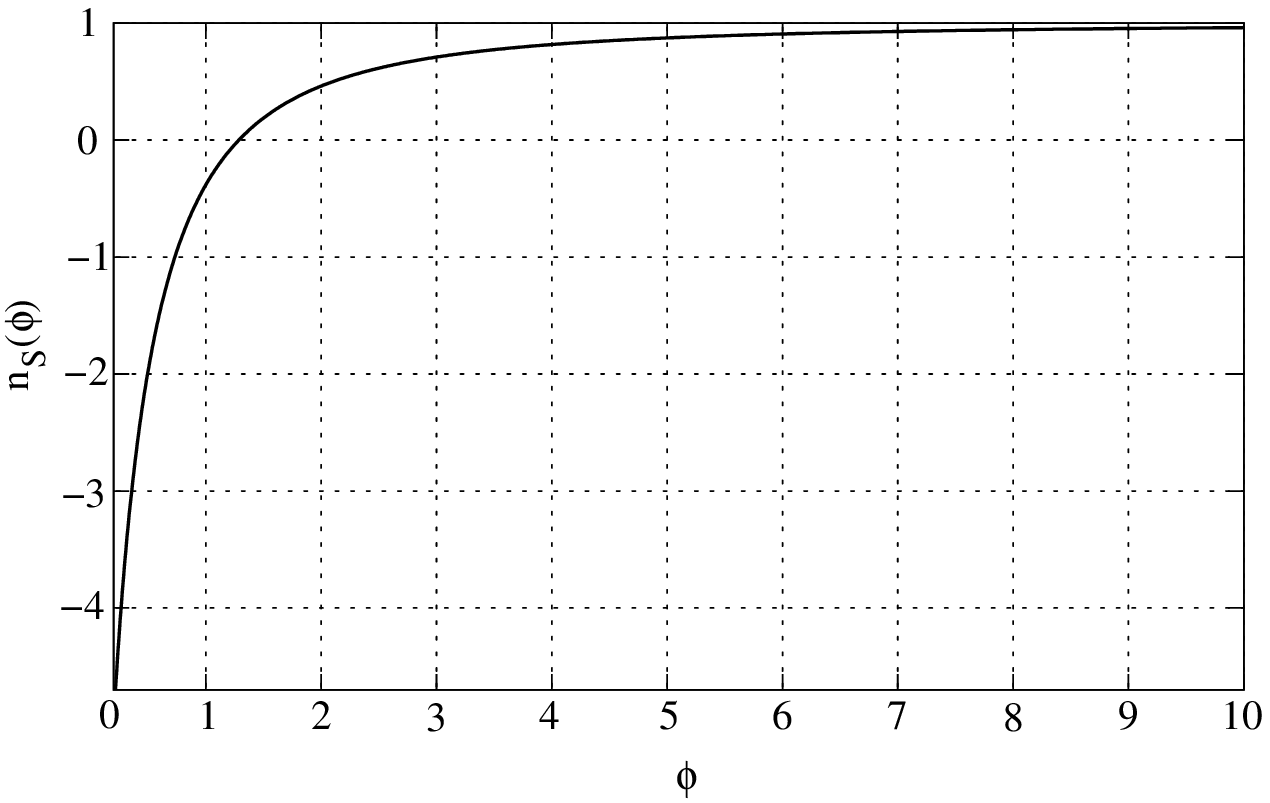, height=7cm}

\epsfig{file=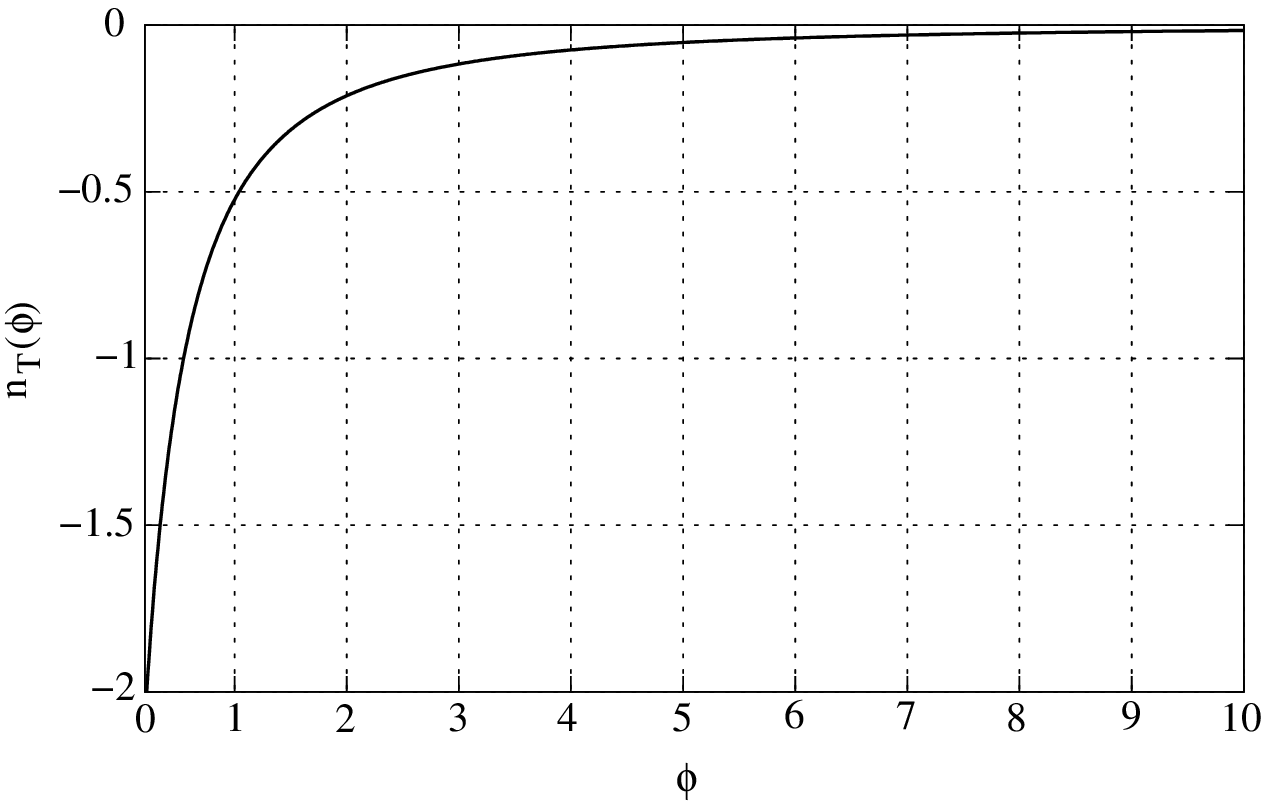, height=7cm}
  \caption{The spectral and tensor indexes  as a function of $\phi$ until the
    end of the inflationary regime.}
  \label{fig:NVLambert}
\end{figure}
\end{center}

Let us now study the behaviour of the spectral and tensor indexes in terms of the
first order slow-roll expansion  \cite{Stewart:1993bc,Liddle:1994dx,Lidsey:1995np}. This first order slow-roll expansion is
sufficient for our present needs. These values 
of the spectral and tensor indexes are 
\begin{eqnarray}
  \label{eq:nst}
  n_S&=& 1- 4(W'(\phi))^2+4W''(\phi)\\
  n_T&=& -4(W'(\phi))^2.
\end{eqnarray}
In Figure \ref{fig:NVLambert} we show the plots for both indexes.
The smallest value of the scalar field is $\phi^*$ in these two plots.
Notice that for large values of $\phi$ we get $n_S\simeq 1$ and $n_T\simeq 0$.
Therefore this novel scalar field model based on the choice of $x(\phi)$ given by equation (\ref{eq:xL}), 
provides another example of a model that yields a perfectly scale-invariant Harrison-Zeldovich spectrum for large field inflation. Interestingly also, the rolling down of the potential brings the model towards a radiation-like behaviour for small values of $\phi$.

\section{Form-invariance map}
\label{sec:FIM}
In this section we address the question: What are the similarities between all
the known solutions? In other words, given that we have shown that the exact
solutions are associated with different choices of the generating function
$x(\phi)$, is it possible to go from one solution to another solution by means
of a transformation $x(\phi) \to \bar{x}(\bar\phi)$? The answer to this question
is related to the classical technique of obtaining solutions of ordinary differential equations  involving  a comparison equation and the  construction of  a map that performs the transformation from a comparison equation to another final equation~\cite{Lakin and Sanchez 1970}.

The underlying question is whether there is a form-invariance transformation
between different solutions, that is an invertible map, which we denote by
$\Psi$, 
that preserves  the form of the equations of motion. In recent articles,
\cite{Chimento:1997uj} and 
 \cite{Chimento:2002gb}, this type of symmetry has been discussed. It was shown
 that the Einstein equations of 
 a flat FRW space-time with a perfect fluid 
 \begin{eqnarray}
   \label{eq:e1}
   \frac{N(N-1)}{2} H^2&=&\rho,\label{NFRied}\\
   \dot\rho +N  H(\rho+p)&=&0, \label{NCons}
 \end{eqnarray}
where $\rho$ is the energy density, $p$ the pressure, and $H =\dot a /a$, admit
a non-trivial map, called  a form-invariance transformation, which transforms 
the quantities $(a,H,\rho)$ to  $(\bar a,\bar 
H,\bar \rho)$, so that the latter quantities satisfy the equations 
\begin{eqnarray}
   \label{eq:e2}
  \frac{N(N-1)}{2} \bar H^2&=&\bar\rho,\\
   \dot{\bar\rho} +N \bar H(\bar\rho+\bar p)&=&0.
 \end{eqnarray}
The $\Psi$ map is given by
\begin{eqnarray}
  \label{eq:MapChimento}
  \Psi_\rho: (\rho,H,p)\longrightarrow \left(F(\rho),
    \left({(F(\rho)}/{\rho}\right)^{1/2}H,-(F(\rho) +
    \left({\rho}/{(F(\rho)}\right)^{1/2}(\rho+p){dF(\rho)}/{d\rho}\right),
\end{eqnarray}
where $F$ is an invertible function, that is,
\begin{eqnarray}
\rho &\to & \bar\rho =  F(\rho) \label{FInv__Chim_rho}\\
H &\to& \bar H = \left({(F(\rho)}/{\rho}\right)^{1/2}H \label{FInv__Chim_Friedman} \\
p &\to& \bar p = -F(\rho) +  \left({\rho}/{(F(\rho)}\right)^{1/2}(\rho+p){dF(\rho)}/{d\rho} \label{FInv__Chim_Conservation}   .
\end{eqnarray}
equation~(\ref{FInv__Chim_rho}) defines the transformation, equation~(\ref{FInv__Chim_Friedman}) guarantees the form-invariance of the Friedmann equation (\ref{NFRied}), and finally equation~(\ref{FInv__Chim_Conservation}) imposes the form-invariance of the conservation equation (\ref{NCons}).

If we now consider perfect fluids with a barotropic equation of state
$p=(\gamma-1)\rho$, where $\gamma$ is the barotropic index, the indexes of both
fluids are related by
\begin{eqnarray}
  \label{eq:gammaChimento}
  \bar\gamma= \left(\frac{\rho}{F(\rho)}\right)^{3/2}\frac{dF(\rho)}{d \rho}\gamma.
\end{eqnarray}

An application of this form-invariance transformation to the case of a
self-interacting scalar field was given in  \cite{Chimento:2003qy} for the
case where  $\bar \rho \propto \rho$. Chimento and Lazkoz have shown that it is
possible to transform standard scalar field cosmologies into phantom
cosmologies. This phantom duality transformation is illustrated by the case of
the exponential potential, which as it is well known, is associated with $\bar
\rho \propto \rho$, and constant $\gamma$, hence inducing a power-law solution
(see equations~(\ref{eq:ffsol1})--(\ref{eq:ffsol3})). 
However,  no general
relation  similar to (\ref{eq:MapChimento}) or (\ref{eq:gammaChimento}) was
given  connecting any pair of self-interacting scalars fields yielding other
types of solutions. 

Here we aim at analysing the general form-invariance dualities that may be established between any scalar field solutions. Our formalism is better suited to this purpose than that of reference \cite{Chimento:2003qy}, since it relies on the
quantities that indeed characterise the scalar field solutions as we have shown in the first part of this paper. The quantities considered in (\ref{eq:MapChimento}) stem from a fluid description which does not separate well the roles played by the
kinetic and potential energies of the scalar field since they are combined in
$\rho_\phi$. Establishing a form-invariance duality requires both the assumption of the transform (\ref{FInv__Chim_rho}) and that of an equation of state for the original model, which then yield the dual model quantities  by means of the Eqs~.(\ref{FInv__Chim_Friedman}) and (\ref{FInv__Chim_Conservation}). In the case of a scalar field model, the method of \cite{Chimento:2003qy} requires that one be able to integrate back from the assumption of an equation of state the corresponding form of the potential, a task which can be cumbersome. In our case the  form fo the potential is readily available, since it follows directly from the specification of $x(\phi)$.
     
We recall that a scalar field can be interpreted as a perfect fluid with the well
known correspondence given by equations (\ref{rhophi}) and (\ref{pphi}),   
and thus it is preferable to use the quantities  $(H,x=\dot\phi/H,\phi)$ instead
of $H$, $\rho$ and $p$. Our aim is to explicitly construct the map
\begin{eqnarray}
  \label{eq:Psi}
  \Psi_\phi: (H,x,\phi)\longrightarrow(\bar H,\bar x,\bar\phi),
\end{eqnarray}
such that the equations (\ref{eq:einstein1})  and (\ref{eq:einstein2}) are
form-invariant under this map.

If we let $H$ be transformed into $\bar H$, the requirement that the scalar
field equation, and hence the energy density conservation, be satisfied becomes    
\begin{eqnarray}
  \label{eq:Psi_define2}
  \frac{\bar x^2}{x^2}&=& \left(\frac{H}{\bar H}\right)^2
\frac{{ d}\bar H}{{ d}H}.
\end{eqnarray}
In obtaining the latter result the Friedmann constraint equation is also assumed
to hold in both frames. Another equivalent condition is also derived from the Raychaudhuri
equation 
\begin{eqnarray}
  \label{eq:Psi_define1}
  \left(\frac{d\bar\phi}{d\phi}\right)^2=\frac{d\bar H}{d H}.
\end{eqnarray}
In addition, from equation (\ref {eq:Hphi})
we also get
\begin{equation}
\label{inv_t}
\frac{d \bar \phi}{\bar x \bar H} = \frac{d \phi}{ x  H},
\end{equation}
which translates the fact that the form invariance transformation preserves the
time variable (\ref{eq:timet}), and which amounts to be the condition that defines the correspondence between the equations of state of the dual models, as  characterized  by $x(\phi)$ and $\bar{x}(\bar\phi)$.  

We establish a form-invariance transformation between any pair of scalar field
solutions by selecting the corresponding generating functions $x(\phi)$ and
$\bar x(\bar \phi)$, deriving the corresponding $H(\phi)$ and $\bar H(\bar\phi)$
functions in accordance to equation (\ref {eq:Hphi}), and plugging them into
equation (\ref{inv_t}) from which we derive the relation
$\bar\phi=\bar\phi(\phi)$. Subsequently, we obtain $\bar H=\bar H(H)$ by using
the conditions (\ref{eq:Psi_define2}) and (\ref{eq:Psi_define1}). 

Notice that by choosing any two  functions $x(\phi)$ and $\bar x(\bar \phi)$ we
are in fact, due to equation (\ref{inv_t}), imposing a relation  between $\bar \phi$ and $\phi$. We will show that by considering $x(\phi)=\lambda$ and $\bar x(\bar
\phi)$ arbitrary it is possible to give an exact solution for this last case in a
 very simple way.

\subsection{Proportional Hubble rates}
\label{sec:prop_H}

Let us consider  the simple case where the Hubble parameters are proportional,
that is, $\bar H=c H$, where $c \in
\C\backslash\{0\}$ is a constant which can be complex. This case can be immediately tackled by resorting to the conditions 
(\ref{eq:Psi_define2}) and (\ref{eq:Psi_define1}) to derive 
\begin{eqnarray}
  \label{eq:alpha1}
    \bar \phi=\pm \sqrt{c}\phi,
\end{eqnarray}
and 
\begin{eqnarray}
  \label{eq:alpha2}
  \bar x= \pm \frac{1}{\sqrt{c}}x.
\end{eqnarray}
so  that the form-invariance map is 
\begin{eqnarray}
  \label{eq:MapOurChim}
  \Psi_c:(H,x,\phi)\longrightarrow (c H, \pm \frac{1}{\sqrt{c}}x,\pm\sqrt{c}\phi),
\end{eqnarray}
which, given equation~(\ref{xgamma}), corresponds to the one given in  \cite{Chimento:2002gb,Chimento:2003qy}. Note that from
\begin{eqnarray}
  a/a_0=\exp\left(\int\frac{d\phi}{x(\phi)}\right),
\end{eqnarray}
one has
\begin{eqnarray}
  \bar a a^{-c}= {\rm const}.
\end{eqnarray}

The previous general equations permit us to review, in a very simple way, the case
analysed by Chimento and Lazkoz from our viewpoint. 
In accordance to our results of
Section \ref{Exact solutions} if we choose the pair
$x(\phi) = \lambda$ and $\bar x=\bar\lambda$, then from equation (\ref{inv_t}) we have  
\begin{equation}
\frac{1}{\bar H_0\bar\lambda}e^{\frac{\bar \lambda}{N-1}\bar\phi}{ d}\bar\phi =
\frac{1}{H_0\lambda}e^{\frac{\lambda}{N-1}\phi}{ d}\phi,
\end{equation}
from which we derive
\begin{equation}
\bar\phi = \frac{\lambda}{\bar\lambda}\phi+ \bar\phi_0
\end{equation}
where 
\begin{equation}
\phi_0=\frac{N-1}{\bar\lambda}\left[\nu+\ln \left(\frac{\bar H_0 \bar\lambda^2}{H_0\lambda^2}\right)\right]
\end{equation}
with $\nu$ being an arbitrary integration constant. This result subsequently implies that
\begin{equation}
\bar H = \left(\frac{\lambda}{\bar\lambda}\right)^2 H.
\end{equation}

We can then distinguish the cases where  the constant of proportionality
$c=(\frac{\lambda}{\bar\lambda})^2$ is equal to $\pm 1$ from those where it takes
some other ratio, and we also distinguish the cases where $c$ is positive from
those where it is negative. (Naturally the case where $c=-1$ requires that one
of the $\lambda$ parameters be imaginary). Since the time variable is the same
in the two 
solutions which are linked by the form-invariance, the cases where $|c|\neq 1$
correspond to power-law solutions associated with different values of the
barotropic-index $\gamma$, and hence with different equations of state. The
identity transformation which obviously preserves the equation of state
corresponds to $c=+1$, i.e., $\bar x=\bar\lambda=\lambda=x$.  The cases where
$c$ is negative are quite interesting since they correspond to transformations
between standard scalar field barotropic solutions and phantom solutions with a
negative kinetic energy. The case when $c=-1$, i.e, $\bar\lambda=\pm i\lambda$,
was discussed in reference  \cite{Chimento:2003qy}, and yields $\bar a(t) =
a^{-1}(t)$  \cite{Gasperini:1992em,Gasperini:2002bn}. When $c<0$, but $c\neq
-1$ we have more general transforms between standard and phantom power-law
solutions such $\bar a a^{-c}=1$.
The associated  phantom form-invariance map
\begin{eqnarray}
  \label{eq:Psi_phatom}
  \Psi_{-1}:(H,x,\phi)\longrightarrow (-H, -i x, i\phi)  .
\end{eqnarray}
gives for any scalar field cosmology its phantom counterpart which is obtained from the $\Psi_{-1}$ map.

At this point it is appropriate to make two remarks. First, if we do not restrict the generating functions $x$ and $\bar x$ to be those leading to power-law behaviours from start, the transform $\bar H =c H$ does not necessarily constrain the scalar field model to the case of the
exponential potential. Second, there is one special application of the proportional Hubble rates transformation this is related to form invariance between $N+1$ and $M+1$ dimensional flat FRW
\cite{Cataldo:2005gb}. Indeed, for $c=\frac{\sqrt{6}}{\sqrt{N(N-1)}}$ one can map the solution
for $(N+1)$-dimensional space-time to a $3$-dimensional one; and for
$c=\frac{\sqrt{M(M-1)}}{\sqrt{N(N-1)}}$ from a $(N+1)$-dimensional space-time to
a $(M+1)$-dimensional one.

\subsection{Transforming a power-law solution to any other solution}
Among all solutions described in Section \ref{Exact solutions} the case of the exponential potential, i.e., $x(\phi)=\lambda$ stands out as particularly interesting and simple. Letting $\bar x(\bar\phi)$ arbitrary it is easy to
build the $\Psi_\phi$ map explicitly. Assuming that one chooses the relation
between $\bar x$ and $x$ to take the form
\begin{eqnarray}
  \int_{\bar\phi_0}^{\bar\phi} \bar x (\xi) d \xi = \lambda \phi,
\end{eqnarray}
it follows that if one invert this relation to obtain $\bar\phi=f(\phi)$, where
 $\tilde \lambda =\lambda/(N-1)$, one gets,
\begin{eqnarray}
  \label{eq:f_phi}
  \bar \phi=f(-\tilde\lambda^{-1}\ln H/H_0).
\end{eqnarray}
A simple calculation then yields
\begin{eqnarray}
  \frac{d\bar\phi}{d\phi}=f'(-\tilde\lambda^{-1}\ln H/H_0),
\end{eqnarray}
and thus
\begin{eqnarray}
  \label{eq:dbarHdH.f}
  \frac{d\bar H}{d H}=\left[f'(-\tilde\lambda^{-1}\ln H/H_0)\right]^2.
\end{eqnarray}
Explicitly this map is given by
\begin{eqnarray}
  \Psi: (H,x,\phi)\longrightarrow \left(
\int dH\left[f'(-\tilde\lambda^{-1}\ln H/H_0)\right]^2,
\frac{x}{f'(\phi)},
f(\phi)
\right).
\end{eqnarray}

Notice that by using the solution for the $x=\lambda$ case (\ref{eq:ffsol2}) one
can construct  an asymptotic  solution for $\bar x$ which can be, in some cases, 
also a exact solution   associated to $\bar x$. This  is given by
\begin{eqnarray}
  \label{eq:barx_sol}
  \bar \phi(t)=f\left(-\frac{N-1}{\lambda}\ln\left(\frac{(N-1)^2}{\lambda^2}(t-t_0)\right)\right).
\end{eqnarray}
In what follows we consider a couple of examples. 

\subsubsection{Power-law solutions to the intermediate inflationary solutions}
In the first case we envisage the possibility of mapping the power-law solutions to the intermediate
inflationary solution  of  Ref.~\cite{Barrow:1990}, that is the map $x=\lambda$ to $\bar
x=\lambda/\bar \phi$. It is straightforward to obtain the  map
\begin{eqnarray}
  \label{eq:PL_II}
  \Psi: (H,x,\phi)\longrightarrow\left(
\frac{\tilde\lambda}{\tilde\lambda-2}H^{(1-2/\tilde\lambda)},
\lambda \exp(-\phi),\exp(\phi)\right).
\end{eqnarray}
Note that $f(\phi)=\exp(\phi)$.

Then it follows that the exact solution associated to $\bar x =\lambda/\bar
\phi$ is given by, using (\ref{eq:barx_sol}),
\begin{eqnarray}
  \label{eq:barx_sol1}
  \bar \phi(t)=\left(\frac{N-1}{\lambda^2}\right)^{-(N-1)/\lambda}\cdot(t-t_0)^{-(N-1)/\lambda},
\end{eqnarray}
with the potential
\begin{eqnarray}
  V(\phi)=A\left(\frac{N(N-1)}{2}-\frac{\lambda^2}{2\phi^2}\right)\phi^{-2\lambda/(N-1)}.
\end{eqnarray}

\subsubsection{Power-law solutions to the new exact solution of subsection \ref{Exact solutions}}

Let us now turn to the case of $x=\lambda$ and $\bar x=-(N-1)W'(\bar\phi)$.
In this case it is straight forward to obtain
\begin{eqnarray}
  f(\phi)=-\frac{\lambda}{N-1}\phi e^{-\frac{\lambda}{N-1}\phi},
\end{eqnarray}
and thus
\begin{eqnarray}
  \Psi: (H,x,\phi)\longrightarrow \left(H^{-H \lambda/(N-2)},\frac{\lambda}{f'(\phi)},f(\phi)\right).
\end{eqnarray}

Then it follows that the exact asymptotic solution associated to $\bar x=-(N-1)W'(\bar
\phi)$ is given by, using (\ref{eq:barx_sol}),
\begin{eqnarray}
  \label{eq:barx_sol2}
  \bar \phi(t)=\frac{(N-1)^2(t-t_0)}{\lambda^2}\ln\left(\frac{(N-1)^2(t-t_0)}{\lambda^2}\right),
\end{eqnarray}
with the potential
\begin{eqnarray}
  V(\phi)=A\left(\frac{N(N-1)}{2}-\frac{(N-1)^2W'(\phi)^2}{2}\right)\phi^{2W(\phi)}.
\end{eqnarray}
Expression (\ref{eq:barx_sol2}) extends the solution (\ref{eq:L_sol_1}) to
the $N\ne 3$ case.

\subsubsection{A non-homogeneous transformation}

In \cite{Parsons:1995kt} Parsons and Barrow investigated a transformation which
also permits to generate an infinite family of solutions for the $k=0$ FRW
scalar field cosmologies for $N=3$. Theirs is a particular class of form-invariance
characterized  by $\bar{H}=\alpha^2H+\beta$, and hence $\bar\phi=\alpha \phi$, where
both $\alpha$ and $\beta$ 
are constants.  Applying the equations (\ref{eq:Psi_define1}) and (\ref{inv_t})
of our procedure, we find 
\begin{equation}
  \label{eq:non-h}
  \bar x = x \frac{\alpha H}{\alpha^2H+\beta},
\end{equation}
so that, after choosing $x(\phi)$ which specifies both $H(\phi)$ and
$V(\phi)$, given by equations (\ref{eq:Vxphi}) and (\ref{eq:Hphi}) derive
successively $\bar H(\bar \phi)$ and $\bar x(\bar \phi)$. 
This class of form-invariance transforms illustrates the implication
of adding a constant to the original Hubble rate $H$, that is of a
non-homogeneous form-invariance transformation.
In 
\cite{Parsons:1995kt} the authors considered a transform  relating  old
and new inflation, and in a subsequent work Barrow, Liddle and Pahud \cite{Barrow:2006dh}
applied the latter invariance to derive a generalization of the intermediate inflation
solution. 

If we  consider the case of $x=\lambda$ and the map (\ref{eq:non-h}), which 
thus reads 
\begin{eqnarray}
  \bar x(\bar \phi)= \frac{\lambda}{\alpha}\frac{e^{-\lambda\bar \phi/(2\alpha)}}{e^{-\lambda \bar \phi/(2\alpha)}+\beta}
\end{eqnarray}
we get the corresponding potential, recovering the result of Ref. \cite{Parsons:1995kt},
\begin{eqnarray}
  \label{eq:non-h_V}
  \bar V(\bar \phi)=\bar A
  \left[3\beta^2+\alpha^2\left(6\beta-\frac{\lambda^2}{2}\right)e^{-\lambda\bar \phi/(2\alpha)}+3\alpha^4e^{-\lambda\bar \phi/(2\alpha)}\right].
\end{eqnarray}

Also for the case $x=\lambda/\phi$ the generating function
  \begin{eqnarray}
  \bar x(\bar \phi)= \lambda^2\frac{\alpha
      (\bar\phi/\alpha)^{-\lambda/2-1}}{\left(\beta
        +\alpha^2(\bar\phi/\alpha)^{-\lambda/2}\right)}
\end{eqnarray}
yields  the potential
\begin{eqnarray}
  \label{eq:non-h_V_2}
  \bar V(\bar \phi)=\bar A\left[3-\lambda^2\frac{\alpha^2
      (\bar\phi/\alpha)^{-\lambda-2}}{\left(\beta
        +\alpha^2(\bar\phi/\alpha)^{-\lambda/2}\right)^2}
  \right](\bar\phi/\alpha)^{-\lambda},
\end{eqnarray}
in accordance to Ref. \cite{Barrow:2006dh}.

\section{Slow-roll, perturbations and form-invariance map}

As a joint application of both Sections \ref{Exact solutions} and
\ref{sec:FIM} we discuss  the solutions associated with the potentials that
preserve the slow-roll approximation; this can be done explicitly by imposing
restrictions on the form-invariance map.

It is known that exponential potentials lead to perturbation spectra that are
exact power laws  \cite{Lucchin:1984yf}. In  \cite{Vallinotto:2003vf} a step
was given for a systematic classification of types of inflationary potentials
that yield a constant scalar perturbation indices. The authors obtain  the solutions
associated with these potentials for the Harrison-Zel'dovich case and to general
power-laws case both  to lower order and to next order slow-roll approximation.

It is possible to think of a infinite hierarchy of expressions for the
perturbation spectra and for spectral indices. Due to the complexity of the
problem, only the first two approximation orders are available in general. To
obtain the restriction on the form-invariance map it suffices to consider the first order
of approximation in the slow-roll parameters. The scalar and tensor indices are given by 
the expressions 
\begin{eqnarray}
  n_S-1&\simeq&-4\epsilon -2\eta,\\
  n_T&\simeq=&-2\epsilon,
\end{eqnarray}
So, using equations (\ref{eq:slow-roll_p}), we get
\begin{eqnarray}
  n_S-1&\simeq&-3x^2+2x',\\
  n_T&\simeq&-x^2.
\end{eqnarray}

Since the slow-roll approximation is the dynamical regime where
\begin{eqnarray}
  \epsilon &\ll& 1,\\
  \eta &\ll& 1,
\end{eqnarray}
it imposes restrictions on $x=x(\phi)$.

Consider  $x$ and  $\bar x$, where $x$ corresponds to a
slow-roll solution, then using the form-invariance map one has
\begin{eqnarray}
  \bar \epsilon &=&\frac{x^2}{2 f'(\phi)^2}, \\
  \bar \eta &=& \eta - \frac{x^2}{2}\left(1+\frac{1}{f'}\right)
  +x'\left(1-\frac{1}{f'^2}\right)
  +  x\frac{f''(\phi)}{f'^2},
\end{eqnarray}
where $f$ is the function defined by equation ~(\ref{eq:f_phi}).
This shows that the form-invariance map preserves the slow-roll
approximation provided that the following condition is satisfied 
\begin{eqnarray}
\left\vert
  x'\left(1-\frac{1}{f'^2}\right)
  - \frac{x^2}{2}\left(1+\frac{1}{f'}\right)
  +  x\frac{f''(\phi)}{f'^2}\right\vert\ll 1.
\end{eqnarray}

\section{Conclusions}

In this work we have presented a unified mechanism that generates exact
solutions of scalar field cosmologies by quadratures.  The procedure investigated
here permits to recover almost all known exact solutions, and shows how one may derive new
solutions. In particular, we have derived one novel solution defined in terms of
the Lambert function.  

The solutions are organised in a classification  which depends on the choice of
a generating function which we have denoted by $x(\phi)$. The choice of the
latter reflects the underlying thermodynamics of the model. Cases in which
$x(\phi)$ differs only by an additive constant correspond to the same potential
$V(\phi)$. Conversely, this shows that the selection of a potential does not
fixes the thermodynamical state of the universe. This is a limitation that must
be faced by the efforts of reconstructing the potential from observations. 

We have also discussed how one can transform solutions from one class, i.e.,
characterized by a given choice of $x(\phi)$ into solutions belonging to other
classes. This type of mappings have been termed form-invariance transformations
in the literature  \cite{Chimento:2003qy}. In the present work we have extended these
transformations to include all sorts of solutions and space dimensions. In
particular we have generalised Chimento and Lazkoz's results on the duality of
standard/phantom solutions of power-law models characterized by exponential
potentials. We have, for instance, shown how one can transform these power-law
solution into intermediate inflationary solutions, or into super-inflationary
solutions either phantom or not.    

We must emphasise that the possibility of establishing a unified procedure to
derive exact scalar field  solutions, and to map different classes of solutions
one into another through form-invariant transformations ultimately stems from
the fact that the field equations are a canonical dissipative system in which
the dissipative term is proportional to the square root of the energy of the
scalar field as revealed by the generalised Klein-Gordon equation and by the
Friedmann constraint equation. 

In forthcoming works we extend the procedure and dualities investigated in the
present work to exact phantom solutions and to solutions of
scalar-tensor gravity theories with a perfect fluid  \cite{JPMTCA_ST}.   
\section*{Acknowledgements}

The authors are thankful to Ana Nunes for many helpful discussions, and to John Barrow 
and Luis Chimento for theirs comments on a earlier version of the present work. Financial
support from the portuguese Foundation for Science and  Technology (FCT) under contract PTDC/FIS/102742/2008 is
gratefully  acknowledged.



\end{document}